\documentclass[12pt]{article}
\usepackage{bezier}
\usepackage{amssymb}
\usepackage{epsfig}

\newcommand{\nc}{\newcommand}
\nc{\be}{\begin{equation}}
\nc{\ee}{\end{equation}}
\nc{\bea}{\begin{eqnarray}}
\nc{\eea}{\end{eqnarray}}
\nc{\bela}{\begin{eqnarray*}}
\nc{\eela}{\end{eqnarray*}}

\nc{\eqn}[1]{{(\ref{#1})}}
\nc{\cA}{{\cal A}}
\nc{\cB}{{\cal B}}
\nc{\cC}{{\cal C}}
\nc{\cD}{{\cal D}}
\nc{\cE}{{\cal E}}
\nc{\cF}{{\cal F}}
\nc{\cG}{{\cal G}}
\nc{\cH}{{\cal H}}
\nc{\cI}{{\cal I}}
\nc{\cJ}{{\cal J}}
\nc{\cK}{{\cal K}}
\nc{\cL}{{\cal L}}
\nc{\cM}{{\cal M}}
\nc{\cN}{{\cal N}}
\nc{\cO}{{\cal O}}
\nc{\cP}{{\cal P}}
\nc{\cQ}{{\cal Q}}
\nc{\cR}{{\cal R}}
\nc{\cS}{{\cal S}}
\nc{\cT}{{\cal T}}
\nc{\cU}{{\cal U}}
\nc{\cV}{{\cal V}}
\nc{\cW}{{\cal W}}
\nc{\cX}{{\cal X}}
\nc{\cY}{{\cal Y}}
\nc{\cZ}{{\cal Z}}

\nc{\simo}[1]{{\stackrel{#1}{\simeq}}}
\nc{\geqo}[1]{{\stackrel{#1}{\geq}}}
\nc{\geo}[1]{{\stackrel{#1}{>}}}
\nc{\guo}[1]{{\stackrel{#1}{\succ}}}
\nc{\rbo}{\raisebox}
\nc{\RR} {\rangle \! \rangle}
\nc{\LL} {\langle \! \langle}
\nc{\rmi}[1]{{\mbox{\small #1}}}
\nc{\eq}{eq.~}
\nc{\nr}[1]{(\ref{#1})}
\nc{\ul}{\underline}
\nc{\mc}{\multicolumn}
\nc{\todo}[1]{\par\noindent{\bf $\rightarrow$ #1}}

\nc{\cu}{{\cal u}}

\title{
  \begin{flushright} {\small $\mbox{HD--THEP--98--02}$}
 \end{flushright}
\vskip 2cm
Hopping Parameter Series Construction \\
for Models with \\ 
Nontrivial Vacuum}

\author{
        Thomas~Reisz\thanks{\tt Supported by a Heisenberg Fellowship,
        \hfill\break $\qquad$
        Email address reisz@thphys.uni-heidelberg.de} 
        \\Institut 
        f\"ur Theoretische Physik, 
        Universit\"at Heidelberg, \\ 
        Philosophenweg 16, 
        D-69120 Heidelberg, Germany}

\begin{document}

\maketitle

\begin{abstract}
Hopping parameter expansions are convergent power series. Under general
conditions they allow for the quantitative investigation of
phase transition and critical behaviour.
The critical information
is encoded in the high order coefficients. Recently, 20th order
computations have become feasible and used
for a large class of lattice field models both in finite and
infinite volume. They have been applied to
quantum spin models and field theories at finite temperature.
The models considered are
subject to a global ${\bf Z}_2$ symmetry or to an even larger
symmetry group such as $O(N)$ with $N\geq 2$.

In this paper we are concerned with the technical details
of series computations to
allow for a nontrivial vacuum expectation value $<\rho(x)>\not=0$,
which is typical for models that break a global ${\bf Z}_2$ symmetry.
Examples are scalar fields coupled to an external field,
or manifestly gauge invariant effective models of Higgs field
condensates in the electroweak theory, even in the
high temperature phase.
A nonvanishing tadpole implies an enormous proliferation of graphs
and limits the graphical series computation
to the 10th order.
To achieve the hopping parameter series to comparable order as in the
${\bf Z}_2$ symmetric case, the graphical expansion is replaced by
an expansion into new algebraic objects called vertex structures.
In this way the 18th order becomes feasible.

\end{abstract}


%
%   SECTION 1
%

\section{Introduction}
 
Linked cluster expansions (LCE) provide a way to construct hopping
parameter or high temperature series of correlation functions
for a large class of lattice models \cite{Wortis,ID}.
In contrast to standard perturbation theory,
the series are convergent power series 
(for a recent proof cf.~\cite{pordt1,pordt2} and references
therein).
Under general conditions the domain of convergence extends to the
phase boundary. The nature of the transition such as its location,
the order and the critical exponents, are encoded in the high
order behaviour of the expansion coefficients.

In the recent past, much effort has been done to prolong the series
for various field models on the lattice \cite{LW1}-\cite{butera}.
Fast and efficient methods have been developed
such that susceptibilities and renormalized
coupling constants are available up
to the order 20 in the hopping parameter.
In particular,
they have been generated for lattice field theories
at finite temperature \cite{thomas1} or quantum spin models.
Particular applications concern the
universality classes of high temperature
transitions for O(N)-symmetric field theories \cite{thomas2}
and Gross-Neveu models,
related to the QCD chiral phase transition \cite{thomas3}.
More subtle questions
concerning the applicability of dimensional reduction have been answered.
The series have been generated and used in finite volume together
with finite size
scaling analysis to resolve the order of phase transitions
\cite{mo1}.

The hopping parameter expansion amounts to a graphical representation
of correlation functions. Appropriate graph classes have to be defined and
generated, and their weights have to be computed.
To make high order computation feasible,
graphs are represented algebraically by incidence matrices. This
representation needs to be one-to-one and fast
\cite{thomas1}.

To achieve high orders, some properties of the model and of the lattice
are taken into account in the course of the graph construction.
Beyond others, it is used that the models considered
are subject to a global ${\bf Z}_2$ symmetry or a symmetry group
including ${\bf Z}_2$ as a subgroup.
This implies that only even correlation functions are non-trivial.

There are interesting models both in statistical mechanics and in
quantum field theory where this global ${\bf Z}_2$ symmetry is
absent on physical grounds or by explicit breaking.
We give two examples.

\begin{itemize}

\item
A scalar field theory on the lattice $\Lambda$, coupled to a nonvanishing
external field,
\be
  S(\Phi) =  -\kappa\;
   \sum_{x,y\in\Lambda}^{\quad\;\prime} \Phi(x) \Phi(y)
   + \sum_{x\in\Lambda} \left( \mu \Phi(x)^2
   + \lambda (\Phi(x)^2)^2 \right)
   + H \sum_{x\in\Lambda} \Phi(x),
\ee
where the prime denotes
nearest neighbour interaction, and $H\not=0$.
Putting the hopping parameter $\kappa$ to its critical value
$\kappa_c(\mu,\lambda,H=0)$ and letting $H\to0$, the
critical exponent $\delta$ is determined by the behaviour
of $<\Phi(x)>$.
Knowledge of its series expansion in $\kappa$ allows for a high
precision measurement of $\delta$.

\item
Effective models in which the fluctuation fields are confined in value
to some subset of the real numbers.
An example is provided by the gauge invariant SU(2) Higgs model.
At finite temperature this model is of great importance for the physics
of the early universe.
The high temperature phase transition is highly non-perturbative
in the region of realistic Higgs masses.
On a finite temperature lattice $\Lambda$, integration of the compact gauge
fields without imposing a gauge fixing
generates an effective model that is described by a partition function
of the form \cite{mpr}
\be \label{intro.ew}
  Z \; = \; \int_0^\infty \prod_{x\in\Lambda}
  \left( d\xi(x) \; \xi(x)^3 \right) \;
  \exp{(-V_{eff}(\xi))} .
\ee
The Higgs field condensates and their correlations are obtained
as correlation functions of the field $\xi(x)$.
Although both the measure and the action $V_{eff}$ are formally invariant
under the global ${\bf Z}_2$ transformation $\xi(x)\to -\xi(x)$,
the field $\xi$ is confined to non-negative values, which implies
that $<\xi(x)> \not= 0$ even in the high temperature phase.

\end{itemize}

In this paper we generalize the linked cluster expansion to models
in which the ${\bf Z}_2$ symmetry is explicitly broken 
or does not exist at all.
There are mainly two generalizations to be included.
First, correlation functions with an odd number of fields
do no more vanish in general.
In turn, vertices are allowed to have odd number of lines attached.
This is easily incorporated into the known algorithm and requires only
a slight generalization, although the number of graphs is
considerably enhanced.
Second, as the more subtle point, the vacuum expectation value
of the field, the tadpole moment, does no more vanish in general,
\be\label{int.tadpole}
    < \rho(x) > \; \not= \; 0.
\ee
This property requires the introduction of new graph classes
with different topological properties.
For instance, one-particle irreducible (1PI) correlation functions,
represented as a sum over 1PI graphs, play an important role
in the course of the expansion. Once (\ref{int.tadpole}) holds, however,
1PI refers only to external line channels.
The necessary modifications will be given here.

Furthermore, the combinatorical complexity is drastically enhanced.
We expect that without further improvement the complexity
to 10th order is already as much as the 20th order before.

To achieve comparable order as in the ${\bf Z}_2$ symmetric case
(order 18-20), we replace where required the concept of graphs by 
the concept of vertex structures.
Vertex structures have already been introduced in \cite{thomas1}.
They were used mainly to avoid a recomputation from scratch of LCE
series if the coupling constants of the model are changed. Using them also
avoids severe roundoff errors due to the finite precision of real
coupling constants.

Here, in addition we use vertex structures to avoid the graphical
construction of the tadpoles and the combinatorical problems 
and computational limitations related to them.
In a way, we can view a vertex structures as an equivalence class
in the set of graphs considered so that
every graph belongs to (precisely) one vertex structure,
and such that

\begin{itemize}
\item the graphical details that are not needed for further
computations are hidden, and
\item equivalence classes are as large as possible,
that is, combine as many graphs as
possible. The number of vertex structures then becomes rather small.
\end{itemize}

Summmation over graphs then is replaced by summation over
vertex structures, with appropriate weights.
We intent to make the set of
vertex structures an algebra with operations
that have their analog for graphs (such as concatenation), but
otherwise are much more efficient and fast.
In this algebra the series of tadpoles and the correlation functions
are constructed.

Once we know this structure it is straightforward to implement
the construction of the series expansions in a language like C or C++.

In Sect.~2 we recapitulate the principles of the linked cluster
expansion in graphical terms. This is done only to the extent required
to understand the necessary modifications
for non-zero vacuum expectation values.
We will show that this is conveniently implemented by a second kind of
vertex renormalization, supplemented by a recurrence relation
for computing the tadpole moments.
In order to obtain comparable efficiency and order as in the symmetric
case, in Sect.~3 the algebra of weighted vertex structures is introduced.
The hopping parameter series construction then is done in this algebra,
generalizing the graphical expansion to a vertex structure
expansion.
A summary is given in Sect.~4.

%
%    SECTION 2
%

\section{The graphical approach to series construction.}

%
%    SUBSECTION 2.1
%

\subsection{Principles}

We start with a short survey of the series construction.
We put particularly emphasis on those points important to
understand the differences of models
with non-vanishing vacuum expectation value.
For many details that are the same as for symmetric models 
and that are not repeated here,
the reader is referred to \cite{thomas1}.

Let $\Lambda$
denote a $D$-dimensional hypercubic lattice. It is assumed to be
of shape $L_0\times L_1\times\cdots\times L_{D-1}$,
with the $L_i\geq 4$ even, finite or infinite, and periodic boundary
conditons imposed in the directions of finite length.
We discuss models described by a partition function of the form
\be \label{series.partition}
   \exp{(W(J,v))} = Z(J,v) = \int_{-\infty}^\infty
      \prod_{x\in\Lambda} d\mu(\rho(x)) \;
      \exp{(-S_{hop}(\rho,v)+\sum_{x\in\Lambda}
      J(x)\rho(x) )},
\ee
where $\rho$ denotes a real valued scalar field, and $J$ are
external sources used to generate the correlation functions.
The action $S_{hop}$ is assumed to be of the form
\be \label{series.action}
   S(\rho,v) =  -\frac{1}{2}\;
   \sum_{x\not=y\in\Lambda} \rho(x) v(x,y) \rho(y).
\ee
For pure nearest neighbour interaction,
\be \label{series.hopping}
   v(x,y) \; = \; \left\{ 
   \begin{array}{r@{\qquad ,\quad} l }
    2\kappa\; & {x=y\pm \widehat\nu \rm\; for\; some \; \nu,} \\
    0 & {\rm otherwise,}
   \end{array} \right.
\ee
with $\widehat\nu$ the unit vector in the positive $\nu$th direction,
$\nu=0,\ldots,D-1$.
We confine attention to this case, although more general pair
interactions are feasible \cite{pordt2}.
The single site measure $d\mu(\rho)$ is appropriately bounded for
the partition function to exist and to be analytic for small 
hopping interaction $v(x,y)$ and small external sources $J(x)$.
For instance, let $I$ be some real interval, finite or infinite,
and $\chi_I$ the characteristic function of $I$,
\be 
   \chi_I(\rho) \; = \; \left\{
   \begin{array}{r@{\qquad ,\quad} l }
    1 \; & {\rho\in I} \\
    0 \; & {\rm otherwise.}
   \end{array} \right.
\ee
Then the hopping parameter series discussed below are
convergent for
\be \label{series.mu}
   d\mu(\rho) \; = \; d\rho \; \chi_I(\rho) \; e^{-V(\rho)},
\ee
with $V(\rho)$ bounded from below by
\be \label{series.vbound}
  V(\rho) \geq c\rho^2 - d \; , \qquad c>0,
\ee
cf. \cite{pordt1,pordt2}.
The two examples given in the introduction correspond to
the single site measures
\be
   d\mu(\rho) = d\rho \; \exp{(-\rho^2-\lambda(\rho^2-1)^2 + H\rho)}
\ee
and
\be
  d\mu(\rho) = d\rho \; \rho \; \chi_{[0,\infty[}(\rho) \;
               \exp{(-V(\rho))}
\ee
(where $\rho=\xi^2$, cf. (\ref{intro.ew})),
with $V(\rho)$ bounded according to (\ref{series.vbound}).
Connected correlation functions are obtained by differentiation of
the generating functional,
\be
  W^{(n)} (x_1,\ldots, x_{n} \vert v) =
   <\rho(x_1) \cdots \rho(x_{n}) >^c
    = \left.
   \frac{\partial^{n}}{\partial J(x_1) \cdots 
   \partial J(x_{n})}
    W(J,v) \right\vert_{J=0}.
\ee
Susceptibilities are defined as zero momentum correlations such as
\be \label{series.suscepts}
  \chi_2 =
   \sum_{x\in\Lambda} < \rho(x) \rho(0) >^c ; \quad
  \mu_2 =  \sum_{x\in\Lambda} (\sum_{i=0}^{D-1} x_i^2)
    \, < \rho(x)\rho(0) >^c .
\ee
Renormalized coupling constants are derived from them in
a straightforward way.

The linked cluster expansion is the Taylor expansion 
of the generating functional $W(J,v)$ with respect to
$v(x,y)$,
\be \label{series.generation}
  W(J,v) = \left. \left( \exp{\sum_{x,y\in\Lambda} v(x,y)
   \frac{\partial}{\partial \widehat v(x,y)}} \right) W(J,\widehat v)
   \right\vert_{\widehat v =0}.
\ee
It generalizes in the obvious way to connected correlation functions.
Multiple derivatives of $W$ with respect to
$v(x,y)$ are managed by the identity \cite{Wortis}
\be \label{series.recursion}
  \frac{\partial W}{\partial v(x,y)} = \frac{1}{2} \left( 
   \frac{\partial^2 W}{\partial J(x) \partial J(y)} +
   \frac{\partial W}{\partial J(x)} \frac{\partial W}{\partial J(y)}
   \right) .
\ee
Correlation functions
become series in the hopping parameter $\kappa$, with coefficients
depending on the toplogy of the lattice and the interaction $\mu$,
Eqn.(\ref{series.mu}).

The evaluation of the Taylor expansion rapidly becomes complex
with increasing order, which is the number of derivatives with
respect to $v(x,y)$.
To manage it one is led in a natural way to a graphical device.
In the present context, a graph is a strucutre
\be
  \Gamma = (\cL_\Gamma,\cB_\Gamma,E_\Gamma,\Phi_\Gamma),
\ee
where $\cL_\Gamma$ and $\cB_\Gamma\not=\emptyset$ are disjoint sets,
the set of internal lines and the set of vertices of $\Gamma$, respectively.
$E_\Gamma$ is a map
\be
  E_\Gamma: \cB_\Gamma  \to \{0,1,2,\ldots\}, \qquad
   u \to E_\Gamma(u),
\ee
that assigns to every vertex $u\in\cB_\Gamma$ the number of external lines
$E_\Gamma(u)$ attached to it. If $E_\Gamma(u)\not=0$, $u$ is called an
external vertex of $\Gamma$.
The number of external lines of $\Gamma$ is given by
$\sum_{u\in\cB_\Gamma} E_\Gamma(u)$.
Finally, $\Phi_\Gamma$ is the incidence relation that assigns internal lines
to their two endpoint vertices. Selflines are excluded.
We consider lines as being unoriented, so $\Phi_\Gamma$
maps onto unoriented pairs of vertices
\be
  \Phi_\Gamma: \cL_\Gamma \to \overline{(\cB_\Gamma\times\cB_\Gamma)} .
\ee
where for $v,w\in\cB_\Gamma$ we have
$\overline{(v,w)}=\overline{(w,v)}$.
For every pair $v,w\in\cB_\Gamma$, $v\not= w$, the number of
common lines of $v$ and $w$ is defined by
\be
    m(v,w) = \Phi_\Gamma^{-1}(\overline{(v,w)}).
\ee
The number of lines attached to $u\in\cB_\Gamma$ is given by
\be
  l(u) := \sum_{w\in\cB_\Gamma} m(u,w) + E_\Gamma(u).
\ee
Connectedness is as usual path connectedness.

Two graphs
\be 
  \Gamma_1 = (\cL_1,\cB_1,E_1,\Phi_1), \;
  \Gamma_2 = (\cL_2,\cB_2,E_2,\Phi_2)
\ee
are called (topologically) equivalent if there are two maps
$\phi_1:\cB_1\to \cB_2$ and $\phi_2:\cL_1\to \cL_2$,
such that
\be
  \Phi_2 \circ \phi_2 = \overline\phi_1 \circ \Phi_1,\qquad
  E_2 \circ \phi_1 = E_1,
\ee
where $\circ$ means decomposition of maps, and
\be
  \overline\phi_1: \overline{\cB_1\times\cB_1} \to
      \overline{\cB_2\times\cB_2}, \qquad
   \overline\phi_1(v,w) = (\phi_1(v),\phi_1(w)).
\ee
A symmetry of a graph
$\Gamma = (\cL,\cB,E,\Phi)$ is a pair of maps
$\phi_1:\cB\to\cB$ and $\phi_2:\cL\to \cL$,
such that
\be \label{lce.25}
  \Phi \circ \phi_2 = \overline\phi_1 \circ \Phi, \qquad
  E \circ \phi_1 = E.
\ee
The number of those maps is called the symmetry number of $\Gamma$.

The set of equivalence classes of connected graphs with $E$ external 
and $L$ internal lines is henceforth denoted by
$\cG_E(L)$, and
\be 
  \cG_E \; := \; \bigcup\limits_{L\geq 0} \; \cG_E(L).
\ee
Further graph classes will be introduced when they are needed below.
Graphical expansions are done with respect to equivalence classes
of graphs. In the following a graph $\Gamma$ is understood to represent
the complete class of graphs equivalent to $\Gamma$
in the sense above.

Connected correlation functions get a representation as a sum over
connected graphs, each graph being endowed with the appropriate weight.
In this description a line of a graph represents a hopping propagator.
For instance, the 2-point susceptibility becomes
\bea \label{series.chi2}
   \chi_2 & = & \sum_{L\geq 0} (2\kappa)^L
   \sum_{\Gamma\in\cG_2(L)} a(\Gamma), \\
   \label{series.agamma}
   a(\Gamma) & = & w(\Gamma) \prod_{u\in\cB_\Gamma}
   \stackrel{\circ}{v}_{l(u)}^c(\mu) ,
\eea
where
\be 
   \stackrel{\circ}{v}_{i}^c(\mu) =
   W^{(i)} (0,\ldots, 0 \vert v=0).
\ee
$w(\Gamma)$ is a rational number that accounts for the various
topological symmetry and lattice imbedding numbers of $\Gamma$.
The couplings $\stackrel{\circ}{v}_{i}^c(\mu)$ are real numbers
and depend only on the local coupling constants of the model,
that is on the single link measure $\mu$.
They are obtained by numerical integration.

The computation of correlation functions proceeds by constructing
more restricted and hence smaller graph classes.
They are then composed of the latter, either analytically
or by concatenation. The first step is the introduction of
1-particle irreducible (1PI) graphs.
A graph is called 1PI if it has the following property:
Remove a single line. The resulting graph then has only one
connected component which has external lines attached.
For instance, the 2-point susceptibility $\chi_2$ becomes
\be
   \chi_2 = \frac{\chi_2^{1PI}}{1-(2\kappa)(2D)\chi_2^{1PI}},
\ee
in graphical terms
%%%%%%%%%%%%%%
% conn --> 1PI 
%%%%%%%%%%%%%%
\be
{
\setlength{\unitlength}{0.7cm}
\begin{picture}(15.0,2.60)
%\put(3.0,1.3){\makebox(0.1,0){$\chi_1 \; = \; <\rho(x)> \; = \; $}}
%
% conn bubble 
%
\put(1.0,1.3){\circle{1.4}}
% vertices
\put(0.3,1.3){\circle*{0.25}}
\put(1.7,1.3){\circle*{0.25}}
% external lines
{\linethickness{1.0pt}
\put(0.3,1.3){\line(-1,0){1.0}}
\put(1.7,1.3){\line(1,0){1.0}}
}
% dashes horizontal
\put(0.3371,1.525){\line(1,0){1.3257}}
\put(0.3371,1.075){\line(1,0){1.3257}}
\put(0.5370,1.825){\line(1,0){0.9260}}
\put(0.5370,0.775){\line(1,0){0.9260}}
%
% =
%
\put(3.2,1.3){\makebox(0.1,0){$=$}}
%
% one 1PI bubble 
%
\put(5.4,1.3){\circle{1.4}}
% vertices
\put(4.7,1.3){\circle*{0.25}}
\put(6.1,1.3){\circle*{0.25}}
% external lines
{\linethickness{1.0pt}
\put(4.7,1.3){\line(-1,0){1.0}}
\put(6.1,1.3){\line(1,0){1.0}}
}
% dashes horizontal
\put(4.7371,1.525){\line(1,0){1.3257}}
\put(4.7371,1.075){\line(1,0){1.3257}}
\put(4.9370,1.825){\line(1,0){0.9260}}
\put(4.9370,0.775){\line(1,0){0.9260}}
% dashes vertical
\put(5.625,0.6371){\line(0,1){1.3257}}
\put(5.175,0.6371){\line(0,1){1.3257}}
\put(5.925,0.8370){\line(0,1){0.9260}}
\put(4.875,0.8370){\line(0,1){0.9260}}
%
% +
%
\put(7.6,1.3){\makebox(0.1,0){$+$}}
%
% two 1PI bubbles 
%
% left bubble
\put(9.8,1.3){\circle{1.4}}
% vertices
\put(9.1,1.3){\circle*{0.25}}
\put(10.5,1.3){\circle*{0.25}}
% external lines
{\linethickness{1.0pt}
\put(9.1,1.3){\line(-1,0){1.0}}
\put(10.5,1.3){\line(1,0){1.0}}
}
% dashes horizontal
\put(9.1371,1.525){\line(1,0){1.3257}}
\put(9.1371,1.075){\line(1,0){1.3257}}
\put(9.3370,1.825){\line(1,0){0.9260}}
\put(9.3370,0.775){\line(1,0){0.9260}}
% dashes vertical
\put(10.025,0.6371){\line(0,1){1.3257}}
\put(9.575,0.6371){\line(0,1){1.3257}}
\put(10.325,0.8370){\line(0,1){0.9260}}
\put(9.275,0.8370){\line(0,1){0.9260}}
%
% right bubble
\put(12.2,1.3){\circle{1.4}}
% vertices
\put(11.5,1.3){\circle*{0.25}}
\put(12.9,1.3){\circle*{0.25}}
% external lines
{\linethickness{1.0pt}
\put(11.5,1.3){\line(-1,0){1.0}}
\put(12.9,1.3){\line(1,0){1.0}}
}
% dashes horizontal
\put(11.5371,1.525){\line(1,0){1.3257}}
\put(11.5371,1.075){\line(1,0){1.3257}}
\put(11.7370,1.825){\line(1,0){0.9260}}
\put(11.7370,0.775){\line(1,0){0.9260}}
% dashes vertical
\put(12.425,0.6371){\line(0,1){1.3257}}
\put(11.975,0.6371){\line(0,1){1.3257}}
\put(12.725,0.8370){\line(0,1){0.9260}}
\put(11.675,0.8370){\line(0,1){0.9260}}
%
% + the other ones
%
\put(14.4,1.3){\makebox(3.0,0){$+\qquad\cdots$}}

\end{picture}
}
\ee
%%%%%%%%%%%%%%
% end conn --> 1PI
%%%%%%%%%%%%%%
with
\be
   \chi_2^{1PI} = \sum_{L\geq 0} (2\kappa)^L
   \sum_{\Gamma\in\cG_2^{1PI}(L)} a(\Gamma),
\ee
that is, $\chi_2^{1PI}$ is composed of 1PI graphs only.

The second necessary step is to compose 1PI graphs
of the even more restricted classes of 1-vertex irreducible
(1VI) graphs and
renormalized moments. This is called vertex renormalization
\cite{Wortis}.
A graph $\Gamma$ is called 1VI if the following condition is satisfied.
Remove an arbitrary vertex $u$ of $\Gamma$, together with the external and
internal lines attached to $u$. We denote the resulting graph by
$\Gamma_u$. Then,
every connected component of $\Gamma_u$ has at least
one external line left attached to one of its vertices.
We write
\be \label{series.s}
  \cS_k(L) = \{ \Gamma\in\cG_k^{\rm 1PI}(L) \;\vert\;
   \Gamma\; \mbox{is 1VI}\; \}
\ee
for the set of graphs that are both 1PI and 1VI.
On the other hand, renormalized moment diagrams are 1PI graphs that have
exactly one external vertex,
\be \label{series.hatq}
  \widehat\cQ_k(L) = \{ \Gamma\in\cG_k^{\rm 1PI}(L) \;\vert\;
   \mbox{$\Gamma$ has 1 and only 1 external vertex} \}.
\ee
The reason for the $\widehat{\quad}$ will become clear in the
next section.
With both of these notions, susceptibilities are represented as
\be 
  \chi_E^{\rm 1PI} = \sum_{L\geq 0} \; (2\kappa)^L\sum_{\Gamma\in\cS_E(L)}
   a(\Gamma) ,
\ee
where the weights $a(\Gamma)$ are computed in the same way as before
by (\ref{series.agamma}),
with the following exception only.
The vertex couplings 
$\stackrel{\circ}{v}_{i}^c(\mu)$ therein are replaced by the
renormalized moments
\be \label{series.2ndren}
  \stackrel{\circ}{v}_{i}^c(\mu) \to v_{i}^c(\kappa,\mu) =
   \sum_{L\geq 0} \; (2\kappa)^L \sum_{\Gamma\in\widehat\cQ_{i}(L)}
   w(\Gamma) \; \prod_{u\in\cB_\Gamma}
   \stackrel{\circ}{v}_{l(u)}^c(\mu) .
\ee
1VI and renormalized moments provide a kind of
product representation of 1PI graphs by
1VI diagrams and renormalized moments. 
In the final end, after having determined all the weights of the graphs,
the coefficients are reorganized to give the series of
$\chi_E^{1PI}$.

Finally, we need to construct all of
$\cS_k$ and $\widehat\cQ_k$. They are generated from a sequence
of appropriate base classes.
Also, in order to do the graph generation on the computer,
we need an algebraic representation of graphs.
This is primarily done by so-called incidence matrices.
The representation
needs to be one-to-one and efficient, that is, fast algorithms are
required to identify graphs by putting them into "canonical"
representations.
These problems are solved such that for O(N) symmetric models
the most important series are available to 20th order in
the hopping parameter.

There are fast algorithms to generate
1-line irreducible (1LI) graphs.
A graph is called 1LI if it cannot be divided in two by cutting
a single line.
For models with vanishing vacuum expectation value
$<\rho>=0$,
a graph is 1LI if and only if it is 1PI, so the notions 1LI and 1PI
are equivalent for them. This concerns the graph classes
$\cS_k$ and $\widehat\cQ_k$.

However, if the expectation value of the field $<\rho>\not= 0$,
the properties of being 1LI and being 1PI are no more the same.
1LI implies 1PI, but not vice versa.
The classes of renormalized moments $\widehat\cQ_k$ become enormously
enhanced.
This problem is addressed in the next subsection.

%
%    SUBSECTION 2.2
%

\subsection{Tadpoles}

Now suppose that the model under consideration is not subject
to a ${\bf Z}_2$ symmetry.
We consider the necessary modifications of the the series
construction.

\begin{itemize}

\item
Correlation functions with an odd number of fields do no more vanish
in general. This implies that higher connected correlation functions
have a more complicated representation in terms of 1PI ones.

\item
Graphs are no more restricted to having vertices only with an even
number of lines attached. "Odd" vertices are allowed now.

\item
The vacuum expectation value of the field does no more vanish in
general,
\be \label{series.tadpole}
   \chi_1 \; = \; <\rho(x)> \; 
   \not= \; 0.
\ee

\end{itemize}

The first two generalizations are easily included. First, the map
between connected and 1PI correlations is done in closed
analytic form. We are still mainly concerned
with 1PI graphs only.
Furthermore, vertices with odd number of lines do not provide
a major problem.
The inclusion of 1LI graphs having vertices with an odd number of lines
attached implies an enormous enhancement of their numbers.
For instance, the graph class $\cS_2(L)$ with number of internal
lines $L\leq18$ grows by about 3 orders of magnitude,
as can be seen from Table 1 below.
However, the algorithms are so efficient that they still work
with this generalization.
They require a minimal modification only,
as long as 1LI graphs are concerned.

A non-vanishing one-point function (\ref{series.tadpole})
provides the major problem.
Graphically we represent this expectation value by the "tadpole",
%%%%%%%%%%%%%%
% tadpole
%%%%%%%%%%%%%%
\be
{
\setlength{\unitlength}{0.7cm}
\begin{picture}(15.0,2.60)
\put(3.0,1.3){\makebox(0.1,0){$\chi_1 \; = \; <\rho(x)> \; = \; $}}
% conn bubble 
\put(7.5,1.3){\circle{1.4}}
% vertex 
\put(6.8,1.3){\circle*{0.25}}
% external line
{\linethickness{1.0pt}
\put(6.8,1.3){\line(-1,0){1.0}}
}
% dashes horizontal
\put(6.8371,1.525){\line(1,0){1.3257}}
\put(6.8371,1.075){\line(1,0){1.3257}}
\put(7.0370,1.825){\line(1,0){0.9260}}
\put(7.0370,0.775){\line(1,0){0.9260}}
\end{picture}
}
\ee
%%%%%%%%%%%%%%
% end tadpole
%%%%%%%%%%%%%%
Now 1PI graphs are no more necessarily 1LI.
In particular, by the very definition, tadpole graphs
are always 1PI,
but almost never 1LI.
To obtain the graphical representation of
1PI n-point functions for $n\geq 2$ we have to

\begin{itemize}

\item
find a way to construct the tadpole graphs, and

\item
attach the tadpoles to 1LI graphs in all possible ways.

\end{itemize}

%%%%%%%%%%%%%%%%%%%%
% tadpole attachment
%%%%%%%%%%%%%%%%%%%%
\be
{
\setlength{\unitlength}{0.7cm}
\begin{picture}(15.0,4.20)
%
%
%
%
% 1PI bubble 
%
\put(3.4,2.1){\circle{1.4}}
% vertices
\put(2.8,2.4606){\circle*{0.25}}
\put(2.8,1.7394){\circle*{0.25}}
% external lines
{\linethickness{1.0pt}
\put(2.8,2.4606){\line(-1,1){0.8}}
\put(2.8,1.7394){\line(-1,-1){0.8}}
}
%
% cdots
%
\put(2.1,2.4){\makebox(0.1,0){$\cdot$}}
\put(2.0,2.1){\makebox(0.1,0){$\cdot$}}
\put(2.1,1.8){\makebox(0.1,0){$\cdot$}}
%
% dashes horizontal
\put(2.7371,2.325){\line(1,0){1.3257}}
\put(2.7371,1.875){\line(1,0){1.3257}}
\put(2.9370,2.625){\line(1,0){0.9260}}
\put(2.9370,1.575){\line(1,0){0.9260}}
% dashes vertical
\put(3.625,1.4371){\line(0,1){1.3257}}
\put(3.175,1.4371){\line(0,1){1.3257}}
\put(3.925,1.6370){\line(0,1){0.9260}}
\put(2.875,1.6370){\line(0,1){0.9260}}
%
% =
%
\put(4.6,2.1){\makebox(3.0,0){$= \; \sum_{m\geq 0} \;\frac{1}{m!}$}}
%
%
% 1LI bubble 
%
\put(9.5,2.1){\circle{1.4}}
% vertices
\put(8.9,2.4606){\circle*{0.25}}
\put(8.9,1.7394){\circle*{0.25}}
\put(10.1,2.4606){\circle*{0.25}}
\put(10.1,1.7394){\circle*{0.25}}
% external lines
{\linethickness{1.0pt}
\put(8.9,2.4606){\line(-1,1){0.8}}
\put(8.9,1.7394){\line(-1,-1){0.8}}
\put(10.1,2.4606){\line(1,1){0.6}}
\put(10.1,1.7394){\line(1,-1){0.6}}
}
%
% cdots
%
\put(8.2,2.4){\makebox(0.1,0){$\cdot$}}
\put(8.1,2.1){\makebox(0.1,0){$\cdot$}}
\put(8.2,1.8){\makebox(0.1,0){$\cdot$}}
%
% dashes horizontal
\put(8.8371,2.325){\line(1,0){1.3257}}
\put(8.8371,1.875){\line(1,0){1.3257}}
\put(9.0370,2.625){\line(1,0){0.9260}}
\put(9.0370,1.575){\line(1,0){0.9260}}
% dashes vertical
\put(9.725,1.4371){\line(0,1){1.3257}}
\put(9.275,1.4371){\line(0,1){1.3257}}
\put(10.025,1.6370){\line(0,1){0.9260}}
\put(8.975,1.6370){\line(0,1){0.9260}}
% dashes diagonal
\put(9.5,2.8){\line(1,-1){0.7}}
\put(8.8,2.1){\line(1,-1){0.7}}
\put(8.86199,2.3880){\line(1,-1){0.9260}}
\put(9.2199,2.73801){\line(1,-1){0.9260}}
\put(9.0050,2.5950){\line(1,-1){0.9899}}
%
% cdots
%
\put(11.0,2.7){\makebox(0.1,0){$\cdot$}}
\put(11.1,2.4){\makebox(0.1,0){$\cdot$}}
\put(11.1,2.1){\makebox(0.1,0){$(m)$}}
\put(11.1,1.8){\makebox(0.1,0){$\cdot$}}
\put(11.0,1.5){\makebox(0.1,0){$\cdot$}}
%
%
% right upper tadpole (as new picture)
%
\put(10.25,3.0){
\setlength{\unitlength}{0.4666cm}
\begin{picture}(5.0,1.5)
%
% 1PI bubble 
%
\put(0.7,0.7){\circle{1.4}}
% vertex
\put(0.2050,0.2050){\circle*{0.35}}
% dashes horizontal
\put(0.0371,0.925){\line(1,0){1.3257}}
\put(0.0371,0.475){\line(1,0){1.3257}}
\put(0.2370,1.225){\line(1,0){0.9260}}
\put(0.2370,0.175){\line(1,0){0.9260}}
\end{picture}
}
%
%
%
% right lower tadpole (as new picture)
%
\put(10.25,0.3){
\setlength{\unitlength}{0.4666cm}
\begin{picture}(5.0,1.5)
%
% 1PI bubble 
%
\put(0.7,0.7){\circle{1.4}}
% vertex
\put(0.2050,1.1950){\circle*{0.35}}
% dashes horizontal
\put(0.0371,0.925){\line(1,0){1.3257}}
\put(0.0371,0.475){\line(1,0){1.3257}}
\put(0.2370,1.225){\line(1,0){0.9260}}
\put(0.2370,0.175){\line(1,0){0.9260}}
\end{picture}
}
\end{picture}
}
\ee
%%%%%%%%%%%%%%%%%%%%
% end tadpole attachment
%%%%%%%%%%%%%%%%%%%%
We can do better.
In the last subsection we have recapitulated the composition of
1PI graphs by 1VI graphs and renormalized moments.
Whereas the 1VI graphs of $\cS_k$ are already 1LI,
as follows from their definiton,
the renormalized moments $\widehat\cQ_k$ are 1PI only.
Hence it is sufficient to do the above tadpole attachment
only to generate $\widehat{Q}_k$.
This amounts to a further vertex renormalization.
We circumvent the construction of 1LI graphs
with large number of external lines, which would be very expensive.

We define the sets of first renormalized moment graphs $\cQ_k(L)$ by
\be \label{series.q}
  \cQ_k(L) = \{ \Gamma\in\cG_k^{\rm 1LI}(L) \;\vert\;
   \mbox{$\Gamma$ has 1 and only 1 external vertex} \},
\ee
and the first renormalized moments themselves by
\be
  \chi_{Q_k} \; = \; \sum_{L\geq 0} (2\kappa)^L
  \sum_{\Gamma\in Q_k(L)}
  w(\Gamma) \; \prod_{u\in\cB_\Gamma}
   \stackrel{\circ}{v}_{l(u)}^c(\mu) .
\ee
The first renormalized moment graphs are 1LI and are constructed as for
${\bf Z}_2$ symmetric models.
The second renormalized moments diagrams $\widehat\cQ_k$,
(\ref{series.hatq}),
are then obtained
by attaching tadpoles to the vertices of the graphs of $\cQ_k$.
Let us assume for the moment that the series representation
of the tadpole moment $\chi_1$ is known, for instance in the form
\be \label{series.chi1}
  \chi_1(\kappa,\mu) \; = \; \sum_{L\geq 0} \; 
  (2\kappa)^L \; \chi_1(L).
\ee
For convenience we introduce the quantities
\bea \label{series.chihat}
   \widehat\chi_1^{(1)}(\kappa,\mu) & = & \sum_{x\in\Lambda}
   v(x,0) \; \chi_1(\kappa,\mu)
   \; = \;
   (2\kappa)\; (2D) \, \chi_1(\kappa,\mu), \nonumber \\
   \widehat\chi_1^{(\nu)}(\kappa,\mu) & = &
   \frac{1}{\nu !} \; (\widehat\chi_1^{(1)}(\kappa,\mu))^\nu \; = \;
   (2\kappa)^\nu \; \frac{(2D)^\nu}{\nu!}
   (\chi_1(\kappa,\mu))^\nu , \quad \nu\geq 2 .
\eea
With the tadpole vertices
$v_{i}^{c,{\rm tadp}}(\kappa,\mu)$
defined by
\be \label{series.1stren}
  v_{i}^{c,{\rm tadp}}(\kappa,\mu) 
  = \stackrel{\circ}{v}_{i}^c(\mu)
  + \sum_{\nu\geq 1} \stackrel{\circ}{v}_{i+\nu}^c(\mu) \;
    \widehat\chi_1^{(\nu)}(\kappa,\mu) ,
\ee
the second vertex renormalization (\ref{series.2ndren})
is replaced by 
\be \label{series.totalren}
  \stackrel{\circ}{v}_{i}^c(\mu) \to v_{i}^c(\kappa,\mu) =
   \sum_{L\geq 0} \; (2\kappa)^L \sum_{\Gamma\in\cQ_{i}(L)}
   \widetilde{a}(\Gamma),
\ee
with
\be \label{series.atwid}
   \widetilde{a}(\Gamma) = w(\Gamma) \prod_{u\in\cB_\Gamma}
   v_{l(u)}^{c,{\rm tadp}}(\kappa,\mu) .
\ee
Finally, after having determined all the weights of the graphs,
the coefficients are reorganized as before to give the series of
$\chi_E^{1PI}$.

We have assumed that the hopping parameter series (\ref{series.chi1})
of the 1-point function is known.
Actually, this series representation is obtained by
(\ref{series.totalren}) itself for $i=1$.
For, we have $v_{1}^c(\kappa,\mu)=\chi_1(\kappa,\mu)$, and
$\widehat\chi_1^{(\nu)}(\kappa,\mu)=O(\kappa^\nu)$.
Hence (\ref{series.chihat}-\ref{series.atwid}) becomes the recursion relation
\be \label{series.chi1recursion}
  \chi_1(\kappa,\mu) =
   \sum_{L\geq 0} \; (2\kappa)^L \sum_{\Gamma\in\cQ_{1}(L)}
   w(\Gamma) \prod_{u\in\cB_\Gamma}
   \left( \stackrel{\circ}{v}_{l(u)}^c(\mu)
    + \sum_{\nu\geq 1}
    \stackrel{\circ}{v}_{l(u)+\nu}^c(\mu) \;
    \widehat\chi_1^{(\nu)}(\kappa,\mu)
   \right) ,
\ee
with initial condition 
$\chi_1(\kappa,\mu) = \stackrel{\circ}{v}_{1}^c(\mu)+O(\kappa)$.
This recursion is solved order by order in $\kappa$.
We need the first renormalized moments $Q_1(L)$ with one external
line, consisting
of 1LI graphs only.

In summary, the generation of the hopping parameter series
of m-point functions to order $M$, say,
according to the discussion above proceeds as follows.

\begin{enumerate}

\item
The single site coupling constants $\stackrel{\circ}{v}_{i}^c(\mu)$
for $1\leq i\leq m+M$ are computed. This amounts to the evaluation of
a sequence of low-dimensional integrals.

\item
Construction of the graphs of $\cS_E(L)$ for $E<=m$ and $L\leq M$,
and of the first renormalized moments $\cQ_E(L)$ for $E+L\leq m+M$,
alltogether being 1LI, and their weight factors. This is done as
for ${\bf Z}_2$ symmetric models.
The only generalization here is that we allow for vertices having an odd
number of lines attached. This is easily included.

\item
The series of the tadpole moment $\chi_1$ are computed by the
recursion relation (\ref{series.chi1recursion}), starting with
$\chi_1(\kappa,\mu) = \stackrel{\circ}{v}_{1}^c(\mu)+O(\kappa)$.
This requires the knowledge of the first renormalized moments $Q_1(L)$
for $L\leq M$, obtained in step 2.

\item
Once the series of $\chi_1$ are known to order $M$, we compute the second
renormalized moments
$v_{i}^c(\kappa,\mu)$ according to
(\ref{series.1stren}-\ref{series.atwid}), by attaching the tadpole
moment in all possible ways to the vertices of the first
renormalized moment graphs
(the first vertex renormalization).
These second renormalized moments then are attached to the vertices of
the the graphs of $\cS_E(L)$ with $E\leq m$ and $L\leq M$
(the second vertex renormalization).
In this way we obtains the series coefficients of the desired 1PI
correlation functions. The final computation of the connected correlations
is done analytically.
Of course each of the steps above involves a sorting of the series
coefficients according to their order in $\kappa$.

\end{enumerate}

%
%    SECTION 3
%

\section{Vertex structures.}

%%%%%%%%%%%%%%%%%%
% table: number of graphs 
%%%%%%%%%%%%%%%%%%

\begin{table}[htb]

\caption{\label{nrgph} The number of inequivalent graphs for the
various 1-line irreducible graph classes defined in Section 2.
For the first renormalized moments $\cQ_E(L)$
this number does not depend on E, the number of external lines.
For comparison we have given also the number of graphs having only
vertices with even number of lines attached. They are denoted by the
superscript ${\;}^{{\rm ev}}$.
Notice that $\cS^{{\rm ev}}_3=\emptyset$.
}
\vspace{0.5cm}

\begin{center}

\begin{tabular}{|r|rr|rr|r|}

\hline
$L$ & $\cQ_2^{\rm ev}(L)$ & $\cQ_2(L)$ &
$\cS_2^{\rm ev}(L)$ & $\cS_2(L)$ & $\cS_3(L)$ \\
[1.0ex] \hline

 0 &      1 & 1 & 1 & 1 & 1 \\
 1 &      0 & 0 & 0 & 0 & 0 \\
 2 &      1 & 1 & 0 & 1 & 1 \\
 3 &      0 & 1 & 1 & 1 & 1 \\
 4 &      4 & 4 & 0 & 4 & 6 \\
 5 &      0 & 6 & 2 & 6 & 12 \\
 6 &     15 & 23 & 3 & 23 & 52 \\
 7 &      0 & 49 & 8 & 53 & 145 \\
 8 &     79 & 174 & 9 & 190 & 567 \\
 9 &      0 & 483 & 40 & 575 & 1954 \\
10 &    439 & 1681 & 68 & 2089 & 7577 \\ [1.0ex] \hline
11 &      0 & 5446 & 247 & 7349 & 28811 \\
12 &   2877 & 19562 & 470 & 27908 & 115290 \\
13 &      0 & 69575 & 1779 & 106666 & 465112 \\
14 &  20507 & 260752 & 3937 & 423687 & 1930270 \\ [1.0ex] \hline
15 &      0 & 989445 & 14801 & 1710920 & 8134945 \\
16 & 161459 & 3874201 & 35509 & 7080691 & 34982052 \\
17 &      0 & 15446928 & 135988 & 29828057 & 152884073 \\
18 &1376794 & 63028306 & 350614 & 128100002 &  \\ \hline
\end{tabular}

\end{center}

\end{table}

%%%%%%%%%%%%%%%%%%
% end table: number of graphs
%%%%%%%%%%%%%%%%%%

%
%   SECTION 3.1
%

\subsection{Motivation}

As we have seen in the last section, the only graphs that are actually
generated and operated on are always 1LI. Their numbers for the
simplest classes are listed in Table 1.
By the procedure outlined at the end of the last section,
all the other moments and correlations are no more obtained as a
graphical representation but directly as hopping parameter series
by the two vertex renormalizations.

Although this construction works in principle, there is need for
improvement. This is for the following reasons.

\begin{itemize}

\item
Beyond exceptional cases,
in most applications the coupling constants are not fixed numbers but
vary over some subsets of the reals,
for instance for the determination of critical surfaces 
and universality domains.
For every point in coupling constant space,
the series construction to 18th order as described above
will take a couple of hours on a Sparc Sun
work station. For many points
this becomes too expensive.

\item
The 1LI graphs generated are needed for each computation and hence
have to be stored. Even in binary coded form this needs a couple of
GByte.

\item
The most severe point is the ill-conditionedness of the composition
procedure. If the weights of the graphs involved would be rational
(as e.g. for the Ising model in zero external field), this problem
is absent. However, whereas the weights $w(\Gamma)$ are rational
numbers, in most cases the vertex couplings
$\stackrel{\circ}{v}_{i}^c(\mu)$ are non-rational.
They are of both signs, of comparable size, and they are computable
to limited precision only.
We expect that roundoff errors rapidly accumulate,
limiting the series computation above to about the 10th order
in the hopping parameter.

\end{itemize}

A solution to this problem is provided by expansion into vertex
structures.
Vertex structures have already been introduced and used for
${\bf Z}_2$ symmetric models \cite{thomas1}.
They provide the possibility to arrange the rational parts of
the weights according to the precsribed concatenation and recursion
relations, keeping track of where and how the vertex couplings
$\stackrel{\circ}{v}_{i}^c(\mu)$ are to be multiplied and added.
The actual multiplication with the vertices is done only in the very end,
when all the other terms have been collected, to obtain the final
series coefficients of the correlation functions. At this stage
then this amounts to a small number of additions and multiplications
(a couple of thousands).
Hence, using operations on vertex structures,
the series construction will become well-conditioned even to large
orders. Also, instead of storing the graphs of $\cS_k$ and $\cQ_k$,
one only needs to keep the vertex structures or their algebraic
representations,
which are both tiny objects and small in number.

%
%   SECTION 3.2
%

\subsection{Definitions and tadpole moment construction}

In order to work with vertex structures and to understand in
which sense they replace graphical manipulations we need to
introduce some basic definitions.

A vertex structure $v$ is an ordered sequence of non-negative
integers
\be \label{vs.def}
   v \; = \; (\nu_i)_{i\in{\bf N}} \; 
   = \; (\nu_1,\nu_2,\dots );
   \qquad \nu_i\in\{0,1,2,\dots\} ,
\ee
with only finitely many $\nu_i\not= 0$.
For $v$ given by (\ref{vs.def}) we call
$v_i \equiv \nu_i$ the $i$th component of $v$,
$s(v) \equiv \max_{v_i\not=0} i$ the size or the length of $v$,
and
$n(v) \equiv \sum_{i\geq1} v_i \,i$ the number of lines of $v$.
The reason for this becomes clear below where a vertex
structure is associated to a graph.
We always have $s(v)\leq n(v)$.
The set of all vertex structures $v$ with $n(v)=n$ is denoted by
$\cV_n$.
We put
\be
   \cV^{(n)} \; = \; \bigcup_{1\leq k \leq n} \cV_k .
\ee
For non-negative integer $k$ we define the vertex structure
$\widehat{k}\in\cV_k$ by
\be
  \widehat{k} = (\nu_i)_{i\in{\bf N}}, \; \mbox{with}\;
  \nu_k=1 \; \mbox{and}\; \nu_i=0 \; \mbox{for} \; i\not=k.
\ee

To a given graph $\Gamma$ a vertex structure $v=v(\Gamma)$ is
associated as follows. We set
\bea \label{vs.vgamma}
   v(\Gamma) & = & (\nu_1,\nu_2,\dots ), \nonumber \\
   \nu_i & = & |\{u\in\cB_\Gamma \; | \; l(u)=i \} |,
\eea
that is, $\nu_i$ is the number of vertices of $\Gamma$ that have
precisely $i$ lines attached.
If $\Gamma$ has E external and L internal lines,
we have $n(v(\Gamma)) = 2L+E$, that is,
$v(\Gamma)\in\cV_{2L+E}$.

For given $\Gamma$, the vertex structure $v(\Gamma)$ is uniquely
defined, but not vice versa.
For given vertex structure $v$ there are many graphs
$\Gamma$ such that $v=v(\Gamma)$. Their number is large but finite.
This allows us to define a very useful equivalence relation 
$\sim_{VS}$ in
the graph classes under consideration.
Hereby, two graphs $\Gamma_1$ and $\Gamma_2$ are equivalent,
$\Gamma_1\sim_{VS}\Gamma_2$,
if $v(\Gamma_1)=v(\Gamma_2)$.
This equivalence relation devides the graph classes into
small sets of equivalence classes, each class being represented by
a single vertex structure.
We will associate appropriate rational weights to vertex structures
and will define operations on them that are equivalent to the
concatenation rules and recursion relations of the last section.

Toward this end we endow $\cup_{n\geq1}\cV^{(n)}$ with the structure
of an associative algebra.
First, for later convenience we define the product of two vertex
structures
$v$ and
$v^{\,\prime}$ by
\be
   v \cdot v^{\,\prime} \; = \; (v_i + v_i^{\,\prime})_{i\in{\bf N}} .
\ee

By some "external" process, such as the graph association above,
a vertex structure gets associated a weight.
A pure, weighted vertex structure is a pair
$(w_v,v)$ of a number $w_v$ of some field $\cF$, for our purpose the
field of rational numbers, and a vertex structure $v$.
For $w_v=1$ we use the notation
$\cP_v=(1 , v)$ and write
$w_v\cdot \cP_v \equiv (w_v,v)$.
A weighted vertex structure then is a finite sum of the form
\be
   (w_v)_{v\in\cV^{(n)}} \; \equiv \;
   \sum_{v\in\cV^{(n)}} w_v \cdot \cP_v .
\ee
This set can be given the structure of an $\cF$ vector space
in the obvious way, denoted by $\cW^{(n)}$.
Furthermore, $\cV_k$ for $1\leq k\leq n$ become sub-vector spaces
of $\cW^{(n)}$, which we denote by $\cW_k$.
Finally, with $\cW_0\equiv\cF$,
\be
    \cW \; = \; \bigoplus_{n\geq 0} \cW_n
\ee
is made a graded algebra with unity by defining for
$v\in\cW_s$, $v^{\,\prime}\in\cW_t$ the product
$\cP_v \cdot \cP_{v^{\,\prime}}\equiv\cP_{v\cdot v^{\,\prime}}\in\cW_{s+t}$,
and by linearity.

It is now easy to translate the series construction into
this algebra.
We examplify this for the computation of the tadpole moment.
First, for the simpler graph classes (the 1LI ones),
summation over graphs is replaced by summation over
associated vertex structures.
This is done by adding the rational weights of all graphs that
belong to the same equivalence class of $\sim_{VS}$.
For instance, 
let us define for $L=0,1,2,\dots$
\be
  \widetilde{Q}_1(L) \; = \; \{ v\in\cV_{2L+1}\; \vert \;
   v(\Gamma)=v \; \mbox{for some $\Gamma\in\cQ_1(L)$} \}.
\ee
For the first renormalized moments with series representation
\be
   \chi_{\cQ_1} \; = \; \sum_{L\geq 0} (2\kappa)^L \; q_1(L),
\ee
the coefficient of $(2\kappa)^L$ is now obtained as
\be \label{vs.vsreps}
  q_1(L) = \sum_{\Gamma\in\cQ_1(L)}
  w(\Gamma) \; \prod_{u\in\cB_\Gamma} \stackrel{\circ}{v}_{l(u)}^c(\mu)
  = \sum_{v\in\widetilde{Q}_1(L)} w_v(Q_1) \prod_{i=1}^{s(v)}
      \left( \stackrel{\circ}{v}_{i}^c(\mu) \right)^{v_i} ,
\ee
where for every vertex structure $v\in\widetilde{Q}_1(L)$
\be
   w_v(Q_1) \; = \; 
   \sum_{\Gamma\in \cQ_1(L)\, {\rm with}\, v(\Gamma)=v }
   w(\Gamma).
\ee
The latter sum consists of a large number of terms, but the
$w(\Gamma)$ are rational numbers.
On the contrary, in (\ref{vs.vsreps}) the various terms are real numbers only,
but their number is very small.
With the identification of
\be \label{vs.vsmap}
   w \cdot \prod_{i=1}^n \left( \stackrel{\circ}{v}_{i}^c(\mu) \right)^{\nu_i}
   \; \simeq \;
    w \cdot \cP_v, \quad v = (\nu_1,\nu_2,\dots ),
\ee
$q_1(L)$ becomes replaced by
\be
   \widetilde{q}_1(L) \; = \;
   \sum_{v\in\widetilde{Q}_1(L)}
   w_v(Q_1) \cdot \cP_v \; \in \; \cW.
\ee
To the lowest orders, we have e.g. on the $4\times\infty^{D-1}$
lattice
\bea
  \widetilde{q}_1(0) & = & 1 \cdot \cP_{(1)} , \nonumber \\
  \widetilde{q}_1(1) & = & 0, \nonumber \\
  \widetilde{q}_1(2) & = &
     D \cdot \cP_{(0,1,1)},             \\
  \widetilde{q}_1(3) & = &
     \frac{D}{3}\cdot \cP_{(0,0,1,1)}, \nonumber \\
  \widetilde{q}_1(4) & = &
     \frac{D}{12}\cdot \cP_{(0,0,0,1,1)}
   + (6D^2-3D+1) \cdot \cP_{(0,3,1)}
   + D^2 \cdot \cP_{(0,1,1,1)} \nonumber \\
   && + \; \frac{D^2}{2} \cdot \cP_{(0,2,0,0,1)}, \nonumber
\eea
where trailing zeros of the vertex structures have been cut off.
The recursion for the series construction of the tadpole moment
$\chi_1(\kappa,\mu)$,
Eqn.~(\ref{series.chi1recursion}), is converted into the
recursion in the algebra $\cW$ as follows.
Let us set
\be
  \widetilde{\chi}_1(L=0) \; = \; \cP_{\widehat{1}},
\ee
which is the translation of the initial condition of
(\ref{series.chi1recursion}).
The analog of (\ref{series.chihat}) becomes
\bea
  \widetilde{\chi}_1^{(1)}(L) & = & 
  2D \cdot \widetilde{\chi}_1(L-1), \quad L\geq 1, \nonumber \\
  \widetilde{\chi}_1^{(\nu)}(L) & = & \frac{1}{\nu} \cdot
    \sum_{L^{\,\prime}=1}^{L-(\nu-1)}
    \widetilde{\chi}_1^{(1)}(L^{\,\prime}) \cdot 
    \widetilde{\chi}_1^{(\nu-1)}(L-L^{\,\prime}),
     \quad 2\leq \nu\leq L .
\eea
The tadpole moment recursion in $\cW$ then reads for $L=1,2,\dots$
\be
  \widetilde\chi_1(L) \; = \; \sum_{L^{\,\prime}=0}^L
  \sum_{v\in\widetilde{Q}_1(L^{\,\prime})} w_v(Q_1) \cdot
   \prod_{i=1}^{l(v)} \prod_{j=1}^{v_i}
              \sum_{L_{ij}=0}^{L-L^{\,\prime}\; \prime}
  \left( \delta_{L_{ij},0} \cdot \cP_{\widehat{i}} +
  \sum_{\nu=1}^{L_{ij}} \cP_{\widehat{i+\nu}} \cdot
  \widetilde{\chi}_1^{(\nu)}(L_{ij}) \right),
\ee
where the prime indicates that the summations are restricted to
$\sum_{ij} L_{ij}=L-L^{\,\prime}$.
The solution of this recurrence relation yields up to some order
$M$, say, the vertex structure representation of
$\widetilde\chi_1(L)$, that is,
\be
   \widetilde\chi_1(L) \; = \;
   \sum_{v\in\cV_{2L+1}} w_v(\chi_1) \cdot \cP_v, \quad
   0\leq L \leq M.
\ee
To the same order as for $\widetilde{q}_1$ above we obtain, again on the
$4\times\infty^{D-1}$ lattice,
\bea
  \widetilde\chi_1(0) & = & 1 \cdot \cP_{(1)} , \nonumber \\
  \widetilde\chi_1(1) & = & 2D \cdot \cP_{(1,1)}, \nonumber \\
  \widetilde\chi_1(2) & = &
     D \cdot \cP_{(0,1,1)}
   + 4 D^2 \cdot \cP_{(1,2)}
   + 2 D^2 \cdot \cP_{(2,0,1)}, \nonumber \\
  \widetilde\chi_1(3) & = &
     \frac{D}{3}\cdot \cP_{(0,0,1,1)}
   + 2 D^2 \cdot \cP_{(1,0,2)}
   + 2 D^2 \cdot \cP_{(1,1,0,1)}
   + 2 D^2 \cdot \cP_{(0,2,1)} \nonumber \\
   &&   + \; 8 D^3 \cdot \cP_{(1,3)}
   + 12 D^3 \cdot \cP_{(2,1,1)}
   + \frac{4D^3}{3} \cdot \cP_{(3,0,0,1)},         \\
  \widetilde\chi_1(4) & = &
     (6D^2-3D+1) \cdot \cP_{(0,3,1)}
   + \frac{D^2}{2}\cdot \cP_{(0,2,0,0,1)}
   + \frac{D}{12} \cdot \cP_{(0,0,0,1,1)}  \nonumber \\
   && + \; \frac{2D^2}{3} \cdot \cP_{(1,0,0,2)}
   + \frac{2D^2}{3} \cdot \cP_{(1,0,1,0,1)}
   + 6 D^3 \cdot \cP_{(2,0,1,1)}
   + 12 D^3 \cdot \cP_{(1,1,2)}     \nonumber \\
   && + \; 8 D^3 \cdot \cP_{(1,2,0,1)}
   + 2 D^3 \cdot \cP_{(2,1,0,0,1)}
   + \frac{5D^2}{3} \cdot \cP_{(0,1,1,1)}
   + 16 D^4 \cdot \cP_{(1,4)}   \nonumber \\
   && + \; 48 D^4 \cdot \cP_{(2,2,1)}
   + \frac{32D^4}{3} \cdot \cP_{(3,1,0,1)}
   + 8 D^4 \cdot \cP_{(3,0,2)}
   + \frac{2D^4}{3} \cdot \cP_{(4,0,0,0,1)}. \nonumber
\eea
The final series representation of the tadpole moment $\chi_1$,
(\ref{series.chi1}), is obtained by the map (\ref{vs.vsmap}),
that is,
\be
   \chi_1(\kappa,\mu) = \sum_{L\geq 0} (2\kappa)^L \; \chi_1(L),
   \qquad
   \chi_1(L) = \sum_{v\in\cV_{2L+1}} w_v(\chi_1) \;
   \prod_{i=1}^{s(v)} 
   \left( \stackrel{\circ}{v}_i^c(\mu) \right)^{v_i}.
\ee
In a similar way as the computation of the tadpole moments, the vertex
renormalization processes are done in terms of vertex structures,
that is in the algebra $\cW$.
This provides a well conditioned algorithm.
To order 18 we have about 5 orders of magnitude less numbers of
vertex structures to deal with than we would have graphs.
In Table~2 we summarize some numbers on vertex structures.

%%%%%%%%%%%%%%%%%%
% table: number of vertex structures
%%%%%%%%%%%%%%%%%%

\begin{table}[htb]
\caption{\label{vertstruct} 
The number of mutually different vertex structures associated with
various graph classes according to (\ref{vs.vgamma}),
used for the computation of the hopping
parameter expansion to order 18.
Compared to the number of graphs as given in Table~1, there are about
5 orders of magnitude less number of vertex structures than graphs.
$\cR_1(L)$ denotes the vertex structures of the tadpole moment
$\chi_1$ to order $L$, that is of the 1-point function
$<\rho(x)>$.
}

\begin{center}

\begin{tabular}{|r|r|rr|r|r|}
\hline
$L$ & $\cQ_1(L)$ & $\cS_2(L)$ &
$\cS_3(L)$ & $\cR_1(L)$ \\
[1.0ex] \hline

 0 &   1 &   1 &   1 &     1  \\
 1 &   0 &   0 &   0 &     1  \\
 2 &   1 &   1 &   1 &     3  \\
 3 &   1 &   1 &   1 &     7  \\
 4 &   4 &   3 &   5 &    15  \\
 5 &   6 &   5 &   8 &    31  \\
 6 &  15 &  12 &  19 &    60  \\
 7 &  22 &  19 &  30 &   110  \\
 8 &  45 &  37 &  55 &   198  \\
 9 &  69 &  58 &  89 &   342  \\
10 & 116 & 103 & 148 &   569  \\ [1.0ex] \hline
11 & 183 & 161 & 229 &   943  \\
12 & 294 & 262 & 364 &  1513  \\
13 & 431 & 399 & 551 &  2377  \\
14 & 677 & 628 & 839 &  3700  \\ [1.0ex] \hline
15 &1003 & 938 &1246 &  5672  \\
16 &1470 &1413 &1838 &  8529  \\
17 &2148 &2068 &2676 & 12745  \\
18 &3119 &3047 &     & 18775  \\ \hline
\end{tabular}

\end{center}

\end{table}

%%%%%%%%%%%%%%%%%%
% end table: number of vertex structures
%%%%%%%%%%%%%%%%%%

%
% SECTION SUMMARY
%
\section{Summary}

We have generalized the linked cluster expansion to models with
nonvanishing vacuum expectation value.
The main emphasis here was on the technical details necessary to
achieve the hopping parameter series of correlation functions
to comparable order as for symmetric models.
The major additional problems that had to be solved are the following ones.

\begin{itemize}

\item
The tadpole moment representing the series expansion of the
vacuum expectation value $<\rho(x)>$ has to be worked out.
It obeys a nonlinear recursion relation that involves 1-line irreducible
moments only.

\item
Computation of connected correlation functions requires the attachment of
tadpole moments to all vertices.
This is conveniently done by including a
second vertex renormalization.

\item
There is a huge proliferation of graphs and in turn of roundoff errors
due to the ill-conditioned real arithmetic for generic 
coupling constants.
At the first sight this limits the computations of the series
to order 10 in the hopping parameter.
The solution of this problem is to
replace the graphical expansion of correlation functions and
moments by an expansion
in terms of weighted vertex structures.
The latter can be given the structure of a graded algebra
$\cW$ with unity
over the rational numbers.
The hopping parameter series construction is done mainly in this algebra.

\end{itemize}

The vertex structures and their weights have to be computed only
once. For the most important moments they are available 
to 18th order. Using them,
for any set of coupling constants (the single site measure
(\ref{series.mu})),
the hopping parameter series themselves are obtained within a couple of
minutes on a Sparc Sun workstation.

We have dicussed real valued fields only, merely for simplicity.
The ideas behind vertex structures are generalized in a 
straightforward way to models with larger symmetry group such as $O(N)$,
broken down e.g. to $O(N-1)$.
This only requires slight changes of equivalence relations
and extensions of the algebra $\cW$.

Various applications are in preparation and will be
presented elsewhere.

%\begin{appendix}
%\input ltriapp1
%\end{appendix}
%%%%%%%%%%%%%%%%%%%%%%%%%%%%%%
%
%  Bibliography
%

%%%%%%%%%%%%%%%%%%%%%%%%%%%

\end{document}